\documentclass{article}
\usepackage{iclr2023_conference,times}
\usepackage{hyperref, url, float}

\usepackage{tikz}
\usepackage{tikz-network}
\usetikzlibrary{shapes.geometric, arrows}
\tikzstyle{startstop} = [rectangle, rounded corners, minimum width=1.5cm, minimum height=0.75cm,text centered, draw=black, fill=red!30]
\tikzstyle{process} = [rectangle, rounded corners, minimum width=1.5cm, minimum height=0.75cm, text centered, draw=black, fill=orange!30]
\tikzstyle{process2} = [rectangle, rounded corners, minimum width=1.5cm, minimum height=0.75cm, text centered, draw=black, fill=blue!30]
\tikzstyle{decision} = [rectangle, rounded corners, minimum width=1.5cm, minimum height=0.75cm, text centered, draw=black, fill=green!30]
\tikzstyle{arrow} = [thick,->,>=stealth]
\tikzstyle{io} = [trapezium, trapezium left angle=70, trapezium right angle=110, minimum width=1.5cm, minimum height=0.75cm, text centered, draw=black, fill=blue!30]

\title{Towards Evology: a Market Ecology Agent-Based Model of US Equity Mutual Funds II}

\author{Aymeric Vi\'{e}
\\
Mathematical Institute, University of Oxford\\
Institute of New Economic Thinking, University of Oxford\\
\texttt{vie@maths.ox.ac.uk} \\
\And
J. Doyne Farmer \\
Mathematical Institute, University of Oxford\\
Institute of New Economic Thinking, University of Oxford\\
Santa Fe Institute \\
}
\iclrfinalcopy 
\begin{document}

\maketitle

\begin{abstract}
Agent-based models (ABMs) are fit to model heterogeneous, interacting systems like financial markets. We present the latest advances in Evology: a heterogeneous, empirically calibrated market ecology agent-based model of the US stock market\footnote{Current build is available open-source at \href{https://github.com/aymericvie/evology}{GitHub}.}. Prices emerge endogenously from the interactions of market participants with diverse investment behaviours and their reactions to fundamentals. This approach allows testing trading strategies while accounting for the interactions of this strategy with other market participants and conditions. Those early results encourage a closer association between ABMs and ML algorithms for testing and optimising investment strategies using machine learning algorithms.

\end{abstract}

\section{Introduction}
\label{introduction}
\paragraph{Motivation} Financial markets are complex adaptive systems featuring a wide diversity of participants. The theory of market ecology \cite{farmer2002market, lebaron2002building, musciotto2018long, lo2019adaptive, levin2021introduction, scholl2021market} borrows concepts from ecology and biology to study financial markets. Trading strategies are analogous to biological species: they exploit market inefficiencies and compete for survival or profit. \cite{scholl2021market} has highlighted the nature of interactions between common trading strategies and the strong density dependence of their returns for stylised trading styles. 

\paragraph{Related work} This research is in the continuity of the vast area of financial agent-based models and market selection with heterogeneous beliefs \cite{blume1992evolution, blume2006if}. For example, several ABMs have recently been introduced for market-making optimisation \cite{spooner2018market}, understanding flash crashes \cite{paulin2019understanding} and providing sophisticated financial architectures for trading training \cite{byrd2019abides}. We attempt to develop the complementary approach of market ecology \cite{farmer2002market, scholl2021market} by focusing on the ecological interactions between the different types of agents and strategies. ABMs are particularly relevant in this perspective because of their inherent emphasis on agent heterogeneity and interactions.

\paragraph{Contribution} We here report on the progress of the Evology project, a large-scale market ecology agent-based modelling, to tackle some pressing questions: using ABMs to forecast market variables, testing trading strategies in financial ABMs to complement backtesting, the influence of market participants (e.g. retail investors, or index funds) on volatility, prices and participants' sizes. Modelling has improved, including more market participants and featuring multiple stocks with historical fundamentals that the ABM agents use to generate endogenous prices. The ABM now features testing of user-specified trading strategies, accounting for heterogeneity and interactions effects that traditional backtesting may miss. Those results further hint at a closer collaboration with ABM-ML for testing and discovering trading strategies in a more realistic artificial stock market. Users of Evology can specify hand-coded functions, and this process can be automated, leading to ML-search for trading rules in this simulation environment. As the model realism improves, notably with research directions outlined in the appendix, the usefulness et relevance of ABM trading strategy testing in Evology will increase.

\section{Model and calibration}
\label{model}
\paragraph{Presentation} Compared to previous versions of Evology and other market ecology models \cite{vie2022towards, vie2022evology, scholl2021market}, this improved agent-based model features a much wider variety of market participants and financial assets. The original participant space focused on active mutual funds now includes domestic/international funds, index funds, ETFs, and retail investors. Those market participants trade shares of multiple stocks whose fundamentals, such as earnings, cash flow, book value, and dividends, are extracted from historical data. 

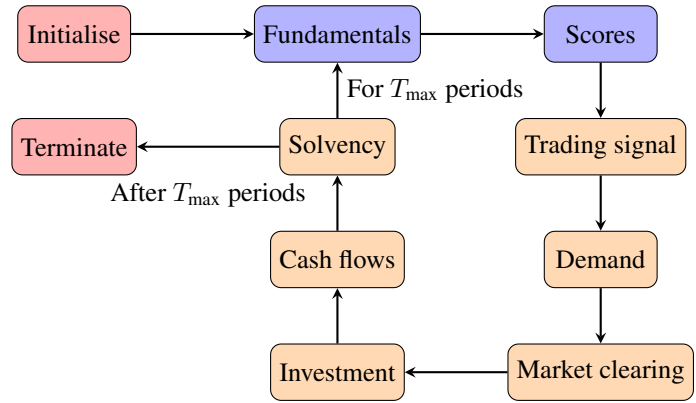
\begin{figure}[ht]
\centering
\begin{tikzpicture}[node distance=1cm]
\node (start) [startstop] {Initialise};
\node (pro0) [process2, right of=start, xshift=2.5cm] {Fundamentals};
\node (pro1) [process2, right of=pro0, xshift=2.5cm] {Scores};
\node (pro2) [process, below of=pro1, yshift=-0.5cm] {Trading signal};
\node (pro3) [process, below of=pro2, yshift=-0.5cm] {Demand};
\node (pro4) [process, below of=pro3, yshift=-0.5cm] {Market clearing};
\node (pro5) [process, left of=pro3, xshift=-2.5cm] {Cash flows};
\node (pro7) [process, below of=pro0, yshift=-0.5cm] {Solvency};
\node (pro6) [process, below of=pro5, yshift=-0.5cm] {Investment};
\node (stop) [startstop, left of=pro7, xshift=-2.5cm] {Terminate};
\draw [arrow] (start) -- (pro0);
\draw [arrow] (pro0) -- (pro1);
\draw [arrow] (pro1) -- (pro2);
\draw [arrow] (pro2) -- (pro3);
\draw [arrow] (pro3) -- (pro4);
\draw [arrow] (pro4) -- (pro6);
\draw [arrow] (pro7) -- node[anchor=west] {For $T_{\text{max}}$ periods} (pro0);
\draw [arrow] (pro6) -- (pro5);
\draw [arrow] (pro5) -- (pro7);
\draw [arrow] (pro7) -- node[anchor=north, yshift = -0.35cm] {After $T_{\text{max}}$ periods} (stop);
\end{tikzpicture}
\caption{Visual summary of the model components. }
\label{model_summary}
\end{figure}

\paragraph{Calibration} In the model \texttt{Initialisation}, we create market participants (Table \ref{participants}) and stocks. At this stage, the model features aggregate market participants: for example, a single agent represents the aggregation of all active Value mutual funds. Stocks, which follow actual stocks with fundamentals data from Compustat IQ Market Intelligence, notably extract their initial prices and number of shares from the data. The activity of the real economy happens exogenously, while the ABM generates prices and wealth dynamics between participants endogenously. \par 

\paragraph{Model steps} Every day, the following actions happen in this market. New \texttt{Fundamentals} are announced by the companies and become public information to market participants; typically, those announcements happen every quarter for each company, and we introduce some delay so that announcements happen uniformly over a 63-day quarter. Based on those fundamentals, the participants can compute stock \texttt{Scores}, which reflect the relative attractivity of stocks for different criteria. Value and Growth scores currently correspond to Morningstar-style classification scales. For example, the Value score includes price to earnings, book, sale, cash flow ratios and dividend yield, while the Growth score features earnings growth, revenue growth etc. From those scores and market information such as market caps/weights, market participants form \texttt{Trading signals}, which act as portfolio weights and determine the wealth participants want to invest in each stock. They formulate \texttt{Demand} functions of the unknown stock price, which depend on the participants' wealth, trading signals, leverage and current stock ownership. For each stock, aggregating participants' demands and finding the root of the resulting aggregate demand leads to the \texttt{Market clearing} price that matches supply with demand. After executing demand orders for the realised prices, participants may receive capital \texttt{Investment} in the form of cash inflows/outflows as a function of their performance. \cite{ha2019misspecifications} analysed N-SAR forms and identified a linear, positive relationship between funds' excess returns and investment flows, suggesting that external investors are chasing returns, confirming earlier results \cite{chevalier1997risk}. Next, cash/stocks pay interest and dividends \texttt{Cash flows}. Before starting the next day of trading, we check the \texttt{Solvency} of participants. The appendix presents additional details on the modelling, calibration results and critical areas of improvement for the ABM.

\begin{table}[t]
\caption{Market participants and behaviours.}
\label{participants}
\begin{center}
\begin{tabular}{ll}
\multicolumn{1}{c}{\bf PARTICIPANT}  &\multicolumn{1}{c}{\bf BEHAVIOR RULE}
\\ \hline \\
Value (active mutual fund)      & Equal weight on top $k$ stocks (by Value score) \\
Blend (active mutual fund)      & Equal weight on top $k/2$ Value \& Growth stocks \\
Growth (active mutual fund)     & Equal weight on top $k$ stocks (by Growth score) \\
Index mutual funds              & Replicate total stock market weights \\
ETFs                            & Assimilated to index mutual funds \\
International mutual funds      & Assimilated to active mutual funds in equal style proportion \\
Retail investors (households)   & Noise trading, weights follow Ornstein Uhlenbeck processes \\
``Strategy fund''               & User-specified \\
\end{tabular}
\end{center}
\end{table}

\section{Results}
\label{results}
\subsection{Snapshot of agent-based model market dynamics}

\begin{figure}[h!]
    \centering
    \includegraphics[width=\textwidth]{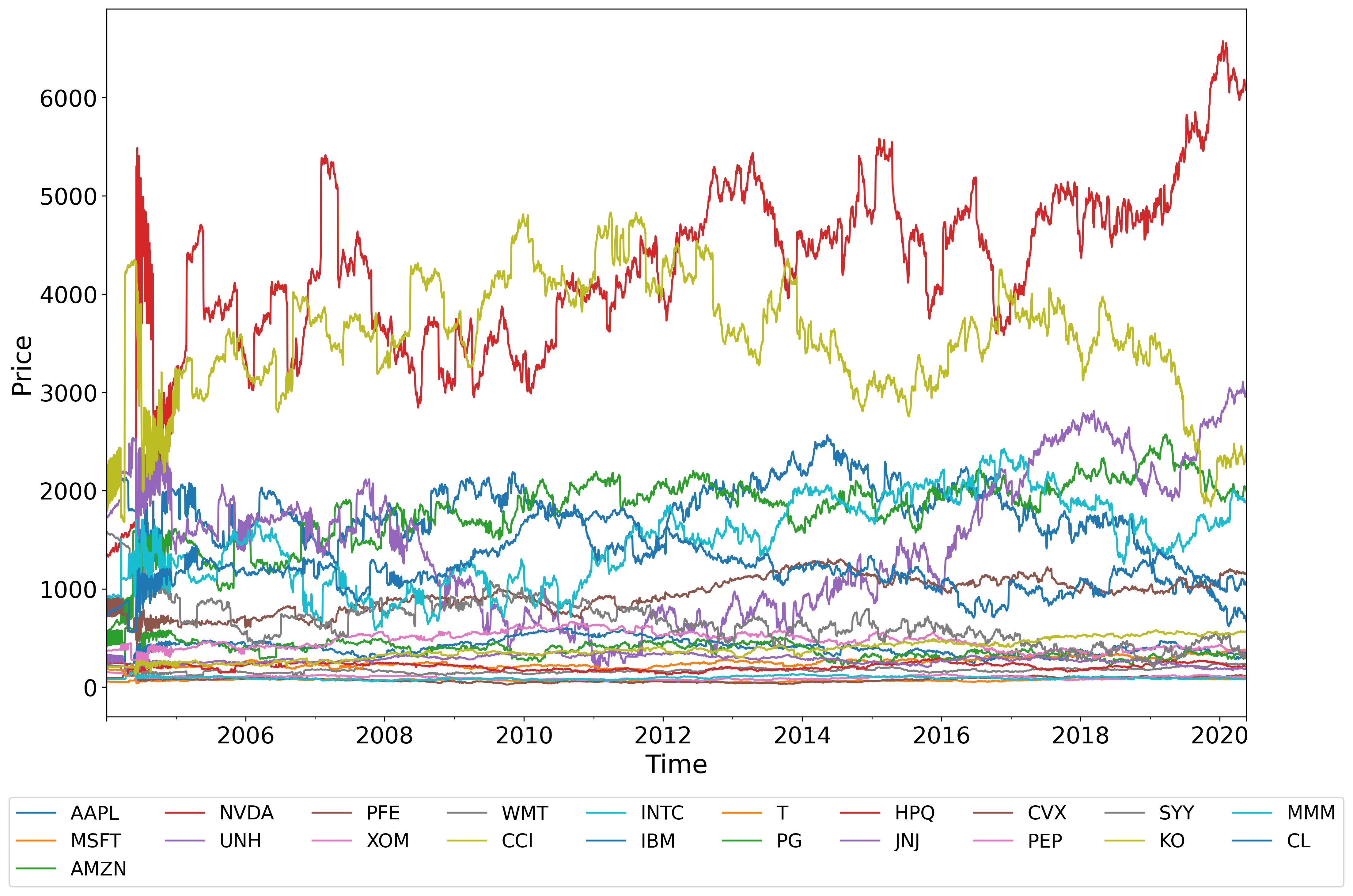}
    \caption{Prices series generated endogenously. }
    \label{prices}
\end{figure}

Evology demonstrates an ability to generate realistic-looking prices by the interactions of different participants with different trading rules and respond to fundamentals announcements by companies (Figure \ref{prices}). The stock market's activity also impacts the wealth and wealth shares of the various market participants. While encouraging, those results highlight many directions for improvement. Simulation runs are stochastic, and prices are currently significantly impacted by the specific Ornstein-Uhlenbeck processes' paths drawn by retail investors. Improvements to Evology should ensure that generated prices match historical prices to validate its forecasting power. Adding more stocks should decrease the average price: at the moment, prices are too high (in the thousands), most likely because all the wealth in the stock market concentrates on only 21 stocks. The early periods of the run are also characterised by high volatility as some stocks move in and out of the top of their score, triggering abrupt changes in the wealth invested in those stocks.

\subsection{Testing trading strategies in Evology}
\begin{figure}[h!]
    \centering
    \includegraphics[width=.5\textwidth]{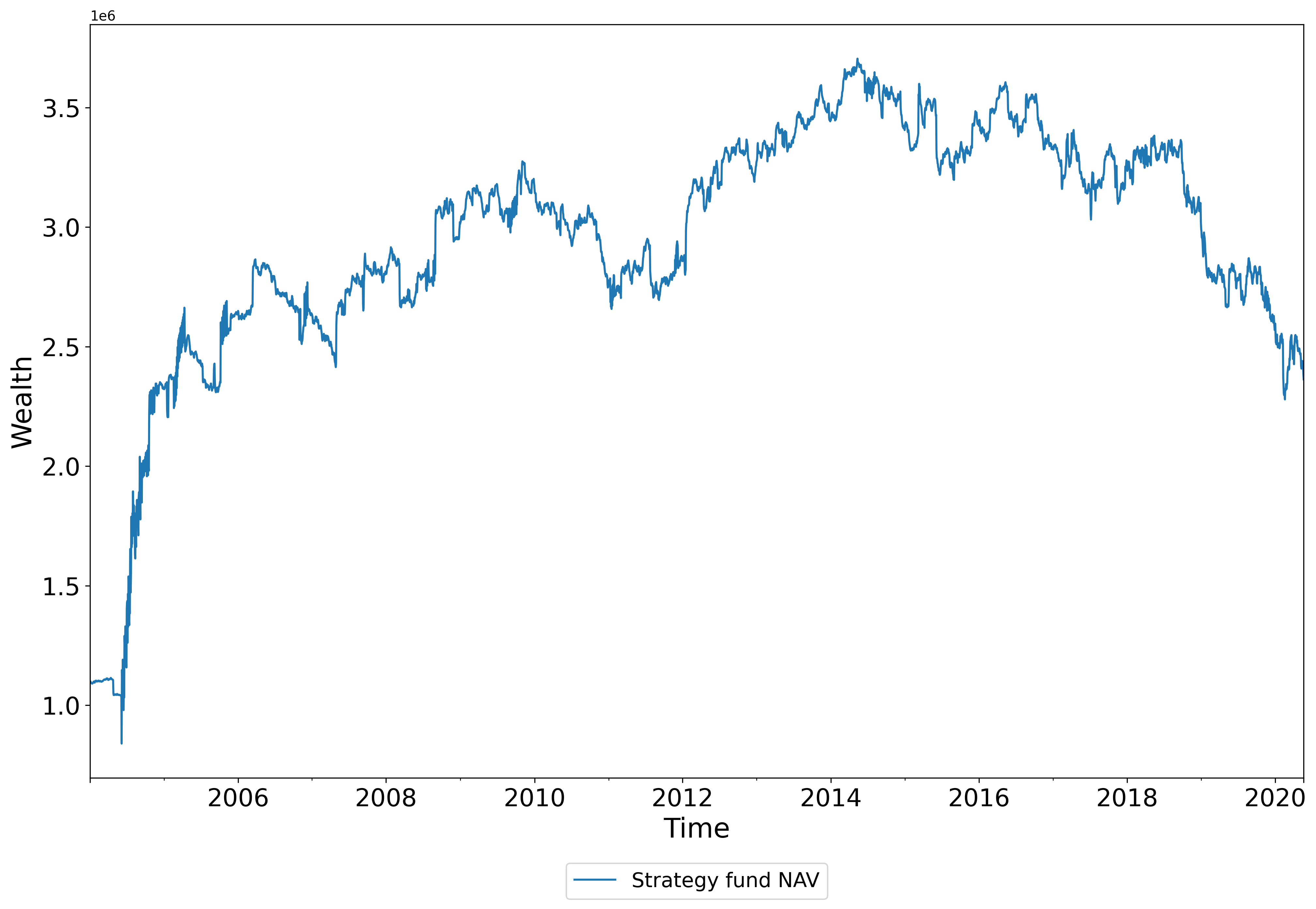}
    \caption{NAV of the user-specified Strategy fund, with strategy "Going 100\% on AAPL"}
    \label{nav_strategy}
\end{figure}

In addition to the market participants in the data mentioned above, the ABM now features a much smaller participant, the ``strategy fund,'' which takes a user-specified strategy and runs it in the market. As a complement to backtesting, this allows testing trading strategies in the ABM and evaluating its return, volatility, and other characteristics while taking into account market impact, interactions with other participants, and crowding effects \cite{farmer2002market}. Using such ABMs to test strategies overcomes the assumption in backtesting that the strategy does not influence the data it uses for evaluation \cite{vie2022towards}. In Figure \ref{nav_strategy}, during the same experiment as Figure \ref{prices}, we show the evolution of the strategy fund NAV when running a straightforward strategy in which the fund invests all its wealth into AAPL shares. The wealth trajectory, in this case, matters less than the ability of the ABM to test a trading strategy while, at the same time, the rest of the market follows its trading rules, an objective introduced in previous research \cite{vie2022towards} and which allows using the ABM as the evaluation part of an ML-driven trading strategy search. Improvements to the ABM realism will translate into gains in real-world relevance for trading strategy testing and greater interest \& challenge for AI/ML algorithms.

\subsubsection*{Acknowledgments}
This publication is based on work supported by the EPSRC Centre for Doctoral Training in Mathematics of Random Systems: Analysis, Modelling and Simulation (EP/S023925/1) and by Baillie Gifford.

\bibliography{main}
\bibliographystyle{iclr2023_conference}

\appendix
\section{Appendix}

\subsection{Modelling details}
\subsubsection{Asset demand}

The funds' daily trading signals are inputs of the \textit{excess demand function} for the asset \cite{scholl2021market}. For each stock, the excess demand function expresses the demand of the fund for the asset as a function of the unknown price $p(t)$. Fund wealth $W$ is the sum of agent cash, the present value of stock shares and liabilities. Our demand function features maximum leverage ${\lambda}$ and strategy aggression $\beta$. Our demand function represents an investor with asset position $S(t)$ and budget $\lambda W(t)$ spending a share $\phi(t)$ of her budget on the asset \cite{poledna2014leverage, scholl2021market}.

\begin{equation}
    D(t, p(t)) =  \beta \phi(t) \frac{\lambda W(t)}{p(t)} - S(t) 
\end{equation}

\subsubsection{Market-clearing}

The \textit{market-clearing} process finds the price for which the sum of the funds' demands equals the fixed asset supply $Q$, demand matching supply \cite{poledna2014leverage}. This is equivalent to the market-clearing procedure of finding the root of the aggregate excess demand function \cite{scholl2021market}. The market-clearing condition for each stock is $\sum_iD_i(t, p(t)) = Q$. While many financial agent-based models use limit-order books (LOBs), our focus here is on long timescales from decades to centuries, while those models focus on intraday dynamics. We do not exclude using LOBs in this model's future but believe market clearing to be a valid approximation for our time horizon.

\subsubsection{Stock scores}

\begin{table}[H]
\caption{Morningstar Growth score. YOY denotes year-over-year growth.}
\label{growth_score}
\begin{center}
\begin{tabular}{ll}
\multicolumn{1}{c}{\bf CRITERION}  &\multicolumn{1}{c}{\bf WEIGHT}
\\ \hline \\
        Long-term projected earnings growth (\%, to now)    & 50\% \\
        Historical earnings growth (\%, YOY)                & 12.5\% \\
        Sales growth (\%, YOY)                              & 12.5\% \\
        Cash flow growth (\%, YOY)                          & 12.5\% \\
        Book value growth (\%, YOY)                         & 12.5\% \\
\end{tabular}
\end{center}
\end{table}

\begin{table}[H]
    \caption{Morningstar Value score}
    \label{value_score}
\begin{center}
\begin{tabular}{ll}
\multicolumn{1}{c}{\bf CRITERION}  &\multicolumn{1}{c}{\bf WEIGHT}
\\ \hline \\
        Price-to-projected earnings  & 50\% \\
        Price-to-book                & 12.5\% \\
        Price-to-sales               & 12.5\% \\
        Price-to-cash flow           & 12.5\% \\
        Dividend yield (\%)          & 12.5\% \\
\end{tabular}
\end{center}
\end{table}

\subsection{Model validation - Empirical stylised facts of asset prices \cite{cont2001empirical}}

Generating the so-called financial ``stylised facts'' is a popular requirement for validating financial ABMs. Our model reproduces the main stylised facts of asset prices \cite{cont2001empirical}. Our log prices display intermittency. The log price returns do not show significant linear autocorrelations past trivial frequencies. Returns show heavy tail distributions with excess kurtosis compared to a normal distribution. We can also reproduce the leverage effect -a negative correlation between price returns and volatility- a positive volume-volatility correlation and slow autocorrelation decay in absolute returns. 

\paragraph{Absence of autocorrelations} Autocorrelations of asset returns should be insignificant except for tiny time scales, in which the microstructure has some impact.

\begin{figure}[H]
    \centering
    \includegraphics[width=0.5\textwidth]{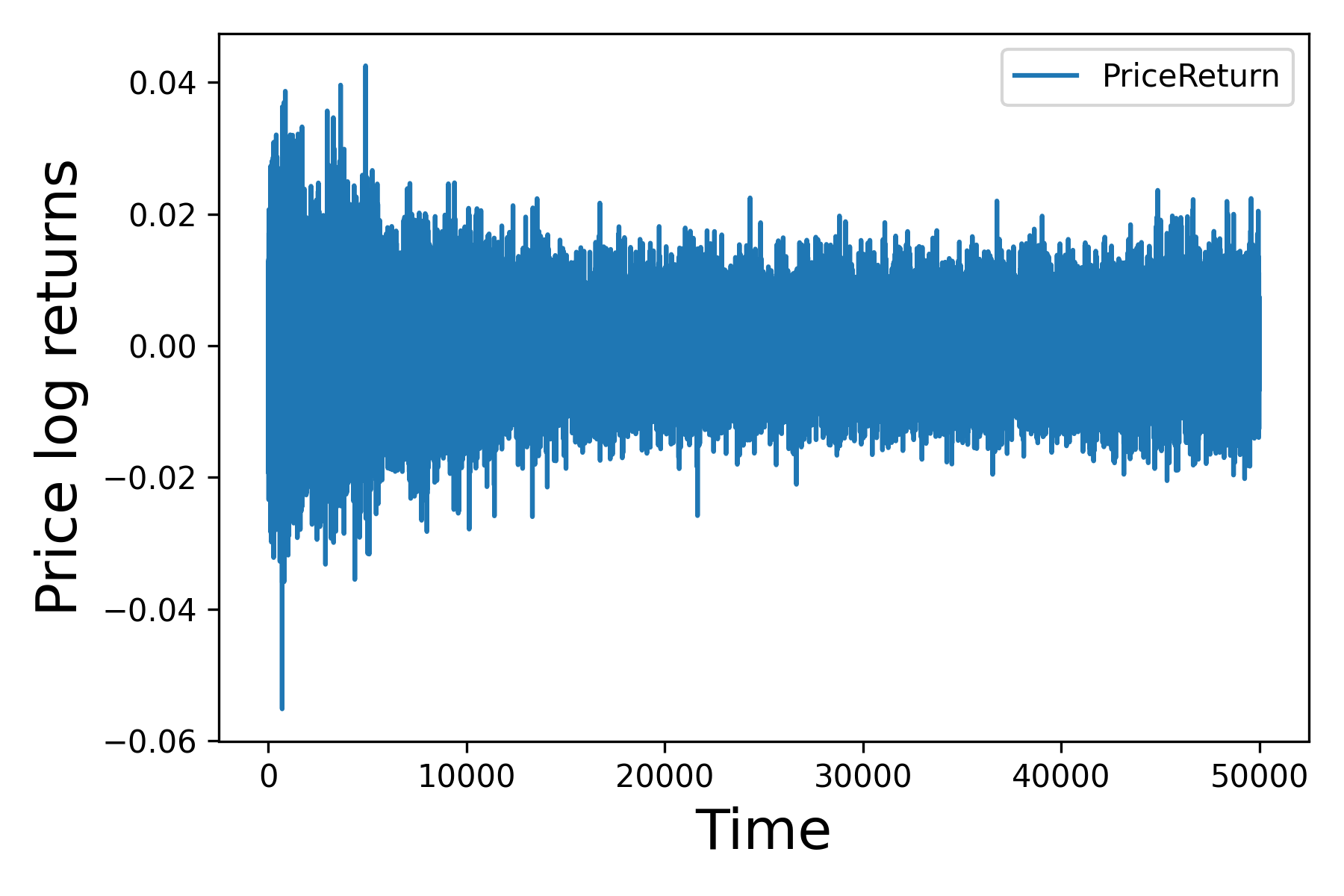}
    \caption{Daily log-returns of the asset price. For a price $p(t)$, the log return at the daily timescale is $r(t) = \ln{p(t)} - \ln{p(t-1)}$.}
    \label{log_returns}
\end{figure}

\begin{figure}[H]
    \centering
    \includegraphics[width=0.5\textwidth]{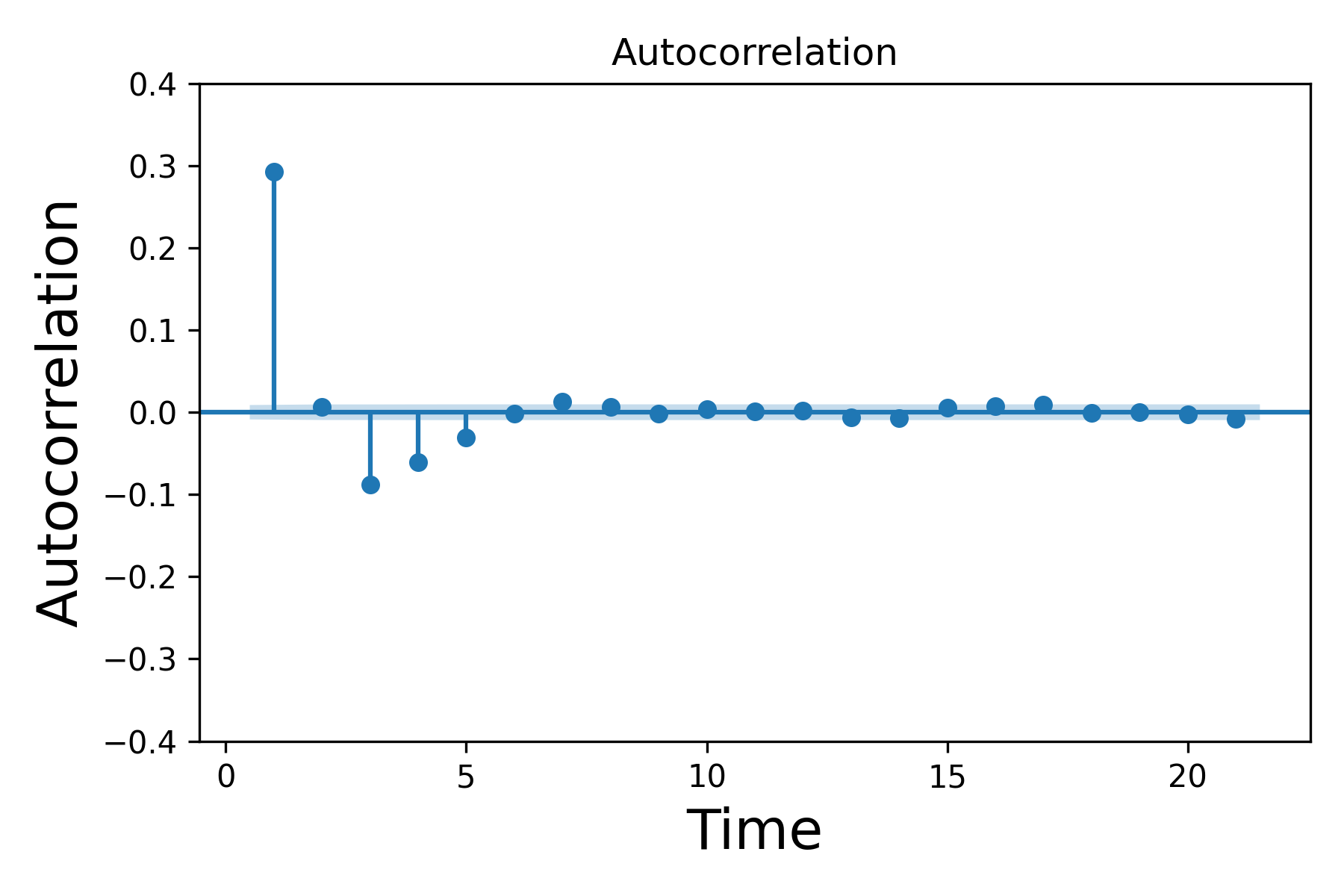}
    \caption{Autocorrelation function of log-returns of the asset price for up to 21 periods. After a short timescale of 5 periods, autocorrelations are not significant from zero, which confirms the absence of autocorrelations.}
    \label{log_returns_acf}
\end{figure}

\paragraph{Heavy tails} The unconditional distribution of returns should display a power-law or a Pareto tail, with finite variance, excluding the normal distribution.

\begin{figure}[H]
    \centering
    \includegraphics[width=0.5\textwidth]{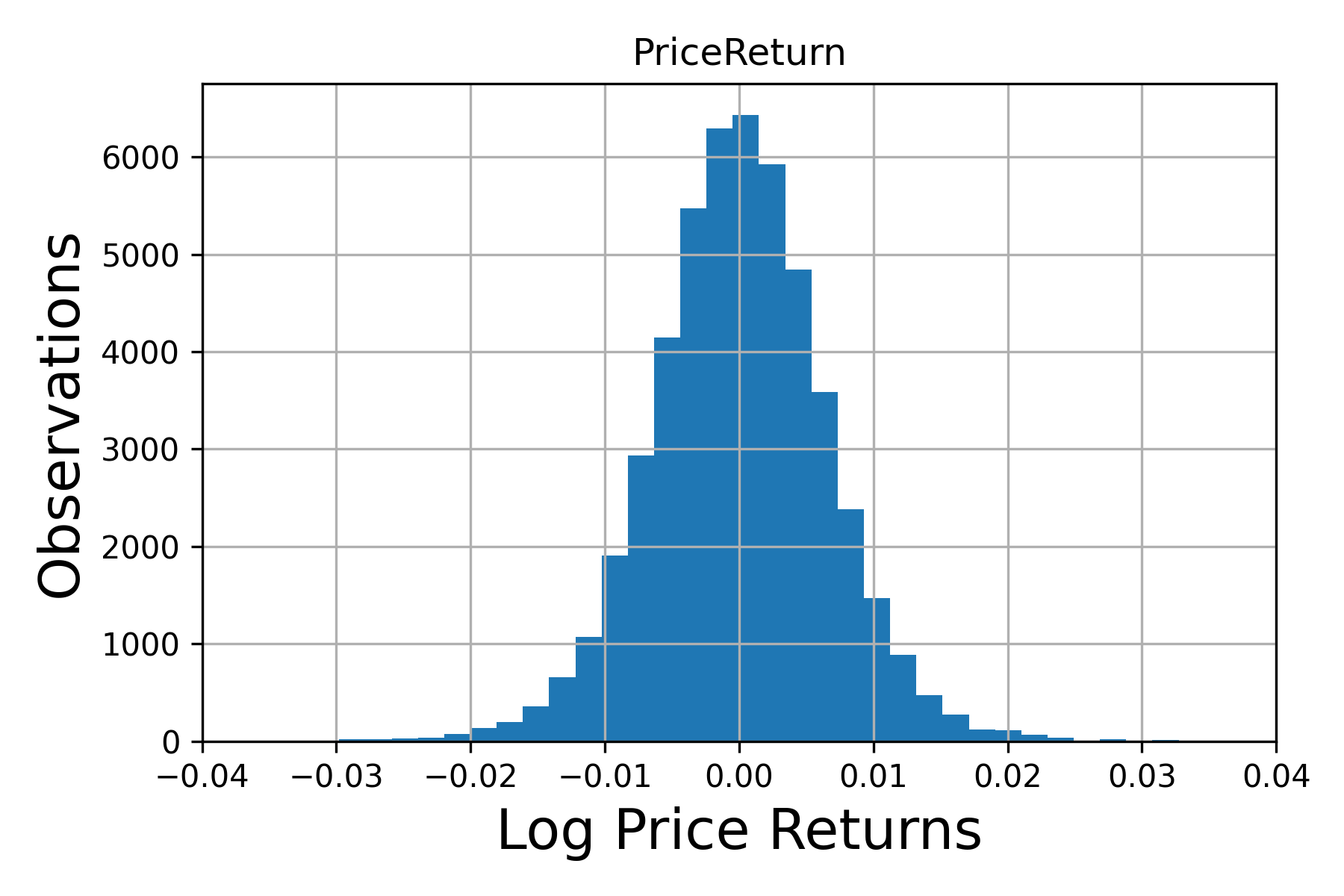}
    \caption{Histogram of the daily log price returns shows heavy tails. The Fisher's (or excess) kurtosis ($\kappa$) value for the series is $1.3$, which is superior to Fisher's kurtosis of the normal distribution, which is equal to $0$. $\kappa = E\left[\left(\frac{X-\mu}{\sigma}\right)^4\right] - 3$.}
    \label{heavy_tail_returns}
\end{figure}

\paragraph{Gain/loss asymmetry} One should observe large drawdowns in stock prices without equally large upward movements. 

\begin{figure}[H]
    \centering
    \includegraphics[width=0.5\textwidth]{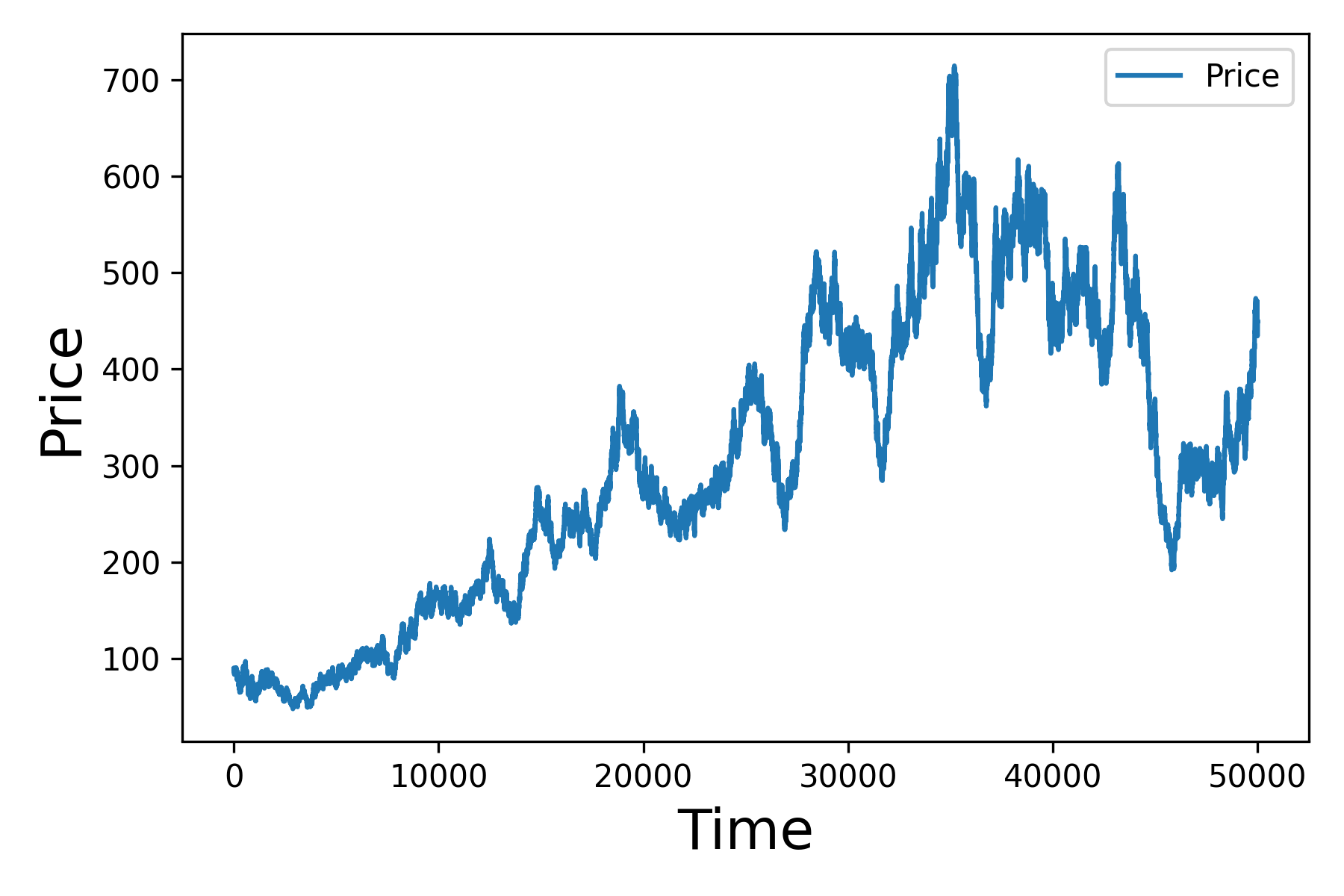}
    \caption{Asset price series of the simulation run. The graphical analysis highlights faster downward than upward price movements: it usually takes more periods to recover from a drawdown than the drawdown duration.}
    \label{regular_price}
\end{figure}

\paragraph{Aggregational Gaussianity} As we increase the timescale for calculating the returns, their distribution should look more and more like a normal distribution.

\begin{figure}[H]
    \centering
    \includegraphics[width=0.5\textwidth]{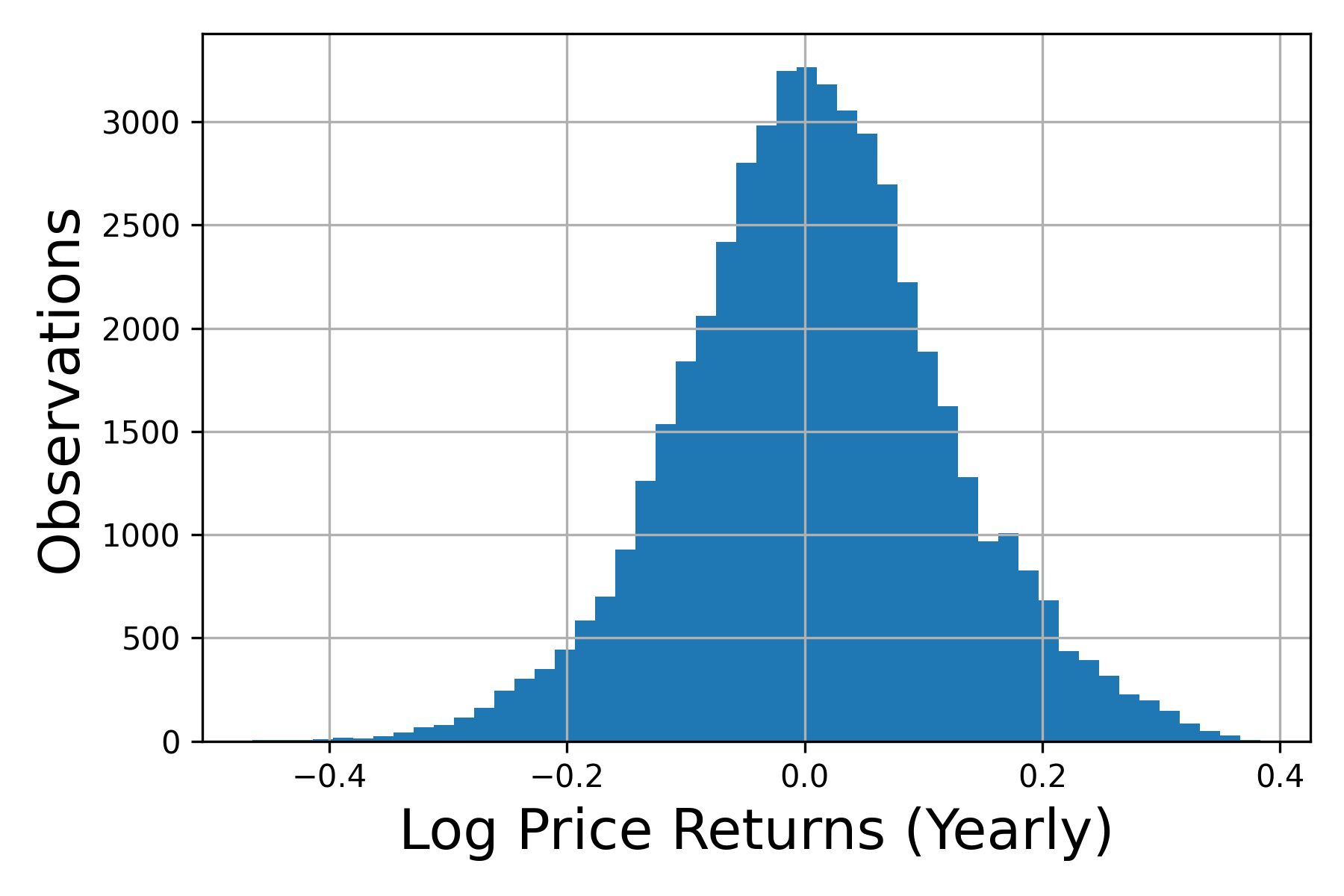}
    \caption{Distribution of yearly log price returns is closer to a normal distribution than the distribution of daily log price returns in Figure \ref{heavy_tail_returns}. Quantitatively, the excess kurtosis for monthly returns is lower, equal to $0.85$, and the excess kurtosis of yearly returns is even lower at $0.32$. Hence, we establish aggregational Gaussianity.}
    \label{distribution_yearly_returns}
\end{figure}

\paragraph{Intermittency} Returns should display high variability, visible by irregular bursts in the time series of volatility estimators.

\begin{figure}[H]
    \centering
    \includegraphics[width=0.5\textwidth]{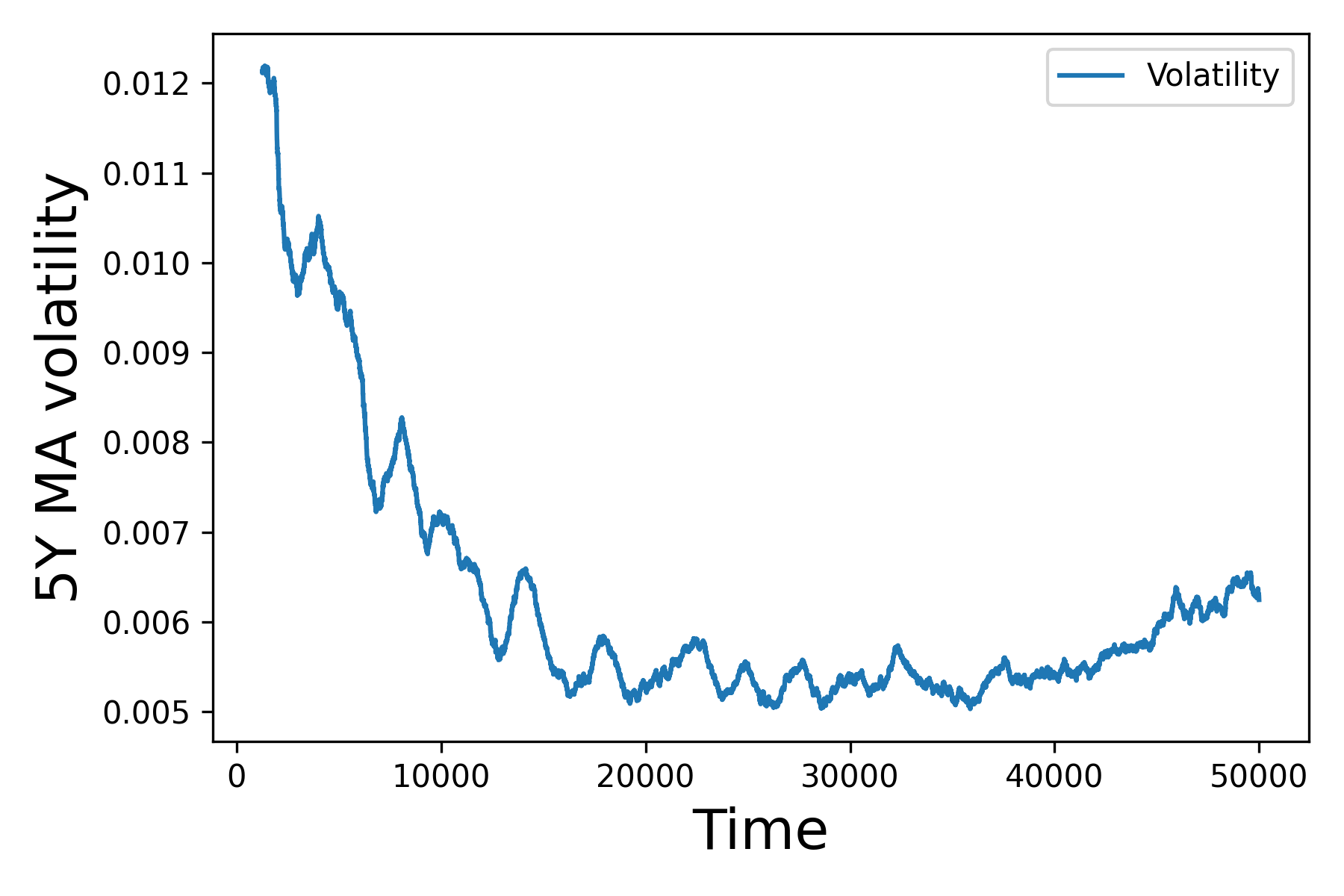}
    \caption{5-year moving average volatility over time show intermittency. We attribute the relatively higher level of volatility in the early steps from the runs to the transient period due to the agent-based model initialisation. However, the irregular bursts in the log-returns of Figure \ref{log_returns} are not obvious, perhaps given the large number of periods represented.}
    \label{intermittency}
\end{figure}

\paragraph{Volatility clustering and slow decay of autocorrelation in absolute returns} Different volatility measures should display a positive autocorrelation over several days, showing that high-volatility events tend to cluster over time. We can observe this volatility clustering from the slow decay of the autocorrelation function of absolute returns.

\begin{figure}[H]
    \centering
    \includegraphics[width=0.5\textwidth]{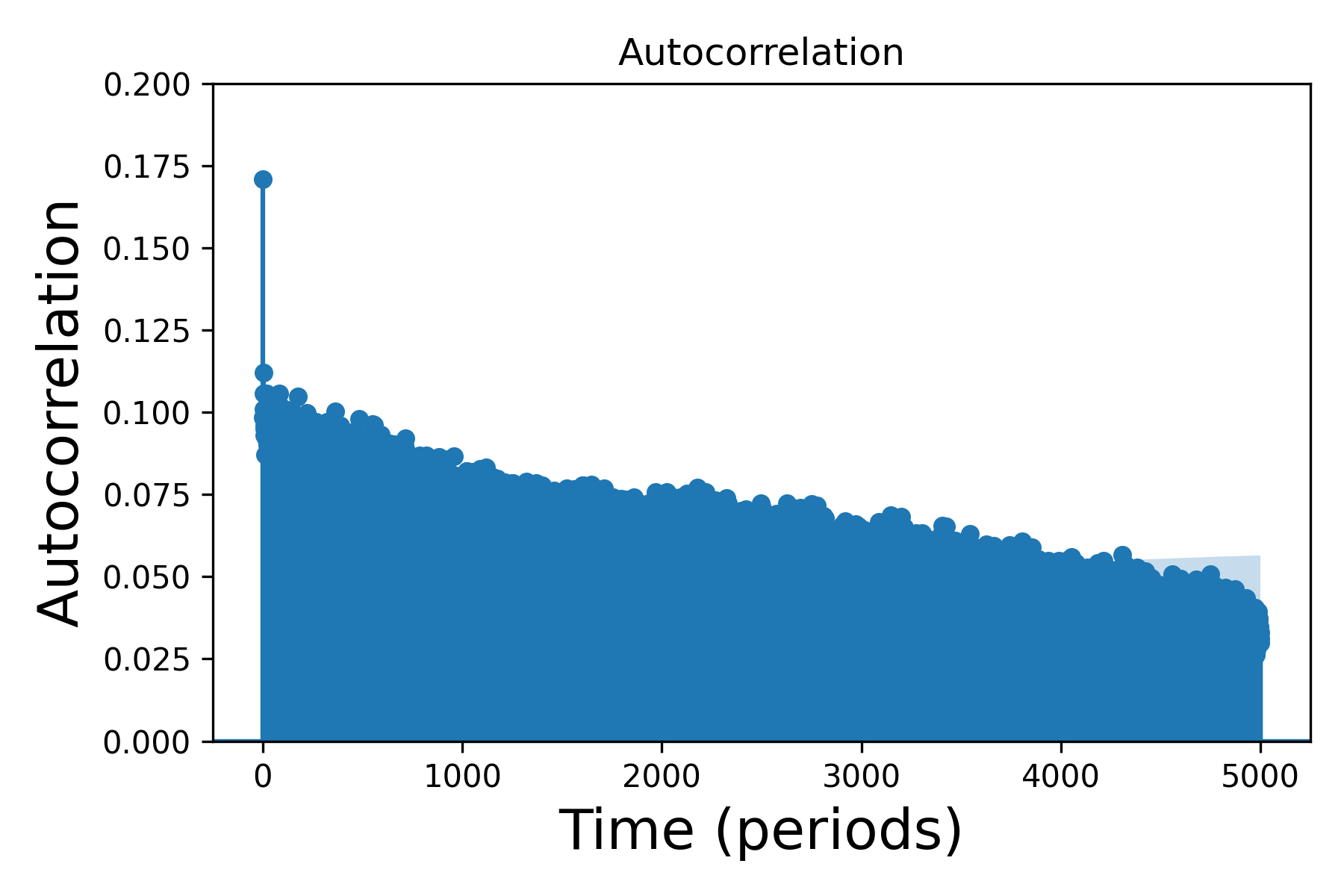}
    \caption{Autocorrelation function of absolute log price returns show long-range dependence and slow decay as a function of time. The autocorrelation remains significantly positive over time lags, showing volatility clustering \cite{cont2007volatility}.}
    \label{volatility_clustering}
\end{figure}

\paragraph{Leverage effect} Measures of asset volatility should negatively correlate with the asset returns. Over our simulation run, this Pearson correlation is negative and equal to $-0.005$.

\paragraph{Volume/volatility correlation} Trading volume should correlate with volatility measures. Over our simulation run, this Pearson correlation is positive and equal to $0.48$. In addition, volume positively correlates with price returns ($0.001$).

\subsection{Improving Evology}
\begin{itemize}
    \item Including missing market participants, in particular pension funds, cash flows and hedge funds. Risk transfer activities also require adding intermediates market participants (market makers, futures/ETF/option arbitrage, wholesalers). 
    \item Modelling interest rates more realistically, from a constant interest rate framework to dynamic interest rates. Historical series could be introduced exogenously and improve interest rate cash flows and add valuation effects to the endogenous prices, i.e. variation in prices not explained by changes in fundamentals but by discount rates.
    \item Improving the calibration is critical for ABMs \cite{paulin2018agent} and asset/mutual funds stylised facts should be respected \cite{cont2001empirical, cont2007volatility, ici22}. In particular, in our case, the calibration of participant sizes at various times should use regulatory filings for institutional participants (13F forms in particular) and other sources for households, e.g. FED household balance sheets. In the future, Evology runs should be capable of covering specific periods (e.g. "01-01-1997" to "31-12-2022") and still be accurately calibrated.
    \item Developing data-driven behaviour rules for the various market participants to better capture the decision-making, at least at an aggregate level, of different investment styles. Better assessing the behaviour and determinants of retail investors' trading activity is particularly important given their importance in the US stock market ownership.
    \item Continuing the expansion of the stock space. From 1 stylised stock \cite{scholl2021market, vie2022towards, vie2022evology}, the current version of Evology now features 21 stocks with historical fundamentals data. As the model progresses into hundreds or even thousands of stocks to become more real-world relevant, market participants' distinctions between small, mid and large cap will be interesting to explore. Some adjustments to the data are also necessary to include stocks that started trading during the simulation period, particularly TSLA and FB/META.
    \item Ensuring the realism of participants' trading rules by checking price returns, participants' returns, turnover ratios, and activity statistics such as liquidity shares.
    \item Improving initialisation of initial stock positions of various participants, finding new data on the stock ownership of those participants. More accurate initial positions will mitigate the volatility around simulation starts as participants reconfigure their portfolios to fit their trading rules.
\end{itemize}

\end{document}